\documentclass[11pt,twoside]{article}
\usepackage{asp2010}

\resetcounters

\begin{document}

\title{Map-making for the Next Generation of Ground-based Submillimeter Instruments}
\author{Gaelen~Marsden$^1$, Adam~Brazier$^2$, Tim~Jenness$^2$,
  Jack~Sayers$^3$, and Douglas~Scott$^1$
\affil{$^1$Department of Physics and Astronomy, University of British
  Columbia, 6224~Agricultural~Road, Vancouver, BC V6T~1Z1, Canada}
\affil{$^2$Department of Astronomy, Cornell University, Ithaca NY,
  14853, USA}
\affil{$^3$California Institute of Technology, 1200 E California Blvd,
  Pasadena, CA 91125, USA}}

\begin{abstract}
  Current ground-based submillimeter instruments (e.g.\ SCUBA-2,
  SHARC-2 and LABOCA) have hundreds to thousands of detectors, sampled
  at tens to hundreds of hertz, generating up to hundreds of gigabytes per
  night. Since noise is correlated between detectors and in time, due
  to atmospheric signals and temperature oscillations, naive
  map-making is not applicable. In addition, the size of the data sets
  makes direct likelihood based inversion techniques intractable. As a
  result, the data reduction approach for most current submm cameras
  is to adopt iterative methods in order to separate noise from sky
  signal, and hence effectively produce astronomical images. We
  investigate how today's map-makers scale to the next generation of
  instruments, which will have tens of thousands of detectors sampled
  at thousands of hertz, leading to data sets of challenging size. We
  propose strategies for reducing such large data sets.
\end{abstract}

\section{Introduction}

The large data volumes expected from upcoming submillimeter (submm)
instruments will pose a challenge for current strategies for
map-making. We begin by describing the map-making problem. We then
discuss solutions to it, and how they apply to current and future
instruments. 

\section{Map-Making}

\subsection{The Map-Making Equation}

\begin{figure}[ht]
\centering
\includegraphics[width=0.5\textwidth]{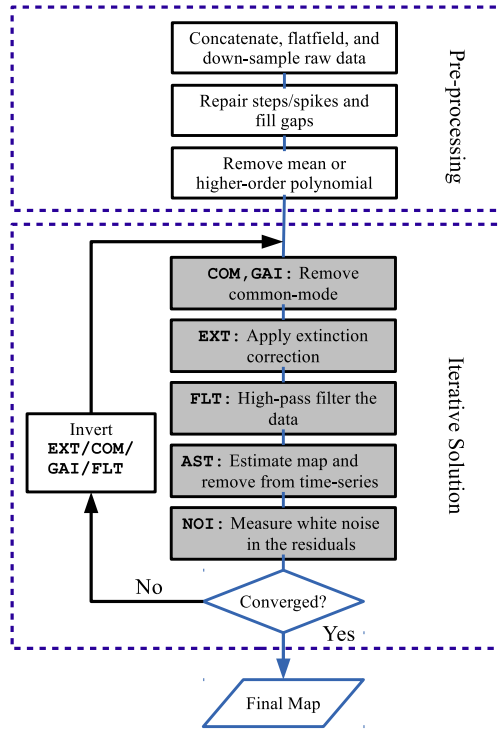}
\caption{ Flow chart reproduced from \citet{2013MNRAS.430.2545C}
  illustrating an example configuration of the iterative map-maker
  SMURF, the SCUBA-2 map-maker. COM, GAI, EXT and FLT are noise
  models. AST represents the sky model and NOI is used to measure the
  white noise levels used in data weighting.
\label{fig:iterative_flowchart}}
\end{figure}

To describe the process of making a map, we begin with the following
model for the data:
\begin{equation}
d_t = A_{ti} s_i + n_t.
\end{equation}
The notation is: $d_t$ is the data, with $t$ indexing both time and
detectors; $A_{ti}$ is the projection operator, associating each time
sample with a map pixel; $s_i$ is the pixelized (and beam-smoothed)
map of the sky, with $i$ indexing the map pixels; and $n_t$ is the
noise in each sample. This is a linear equation, with the
least-squares solution (dropping the indices):
\begin{equation}
\label{eqn:mmsoln}
\mathbf{s} = (\mathbf{A}^\mathrm{T} \mathbf{N}^{-1} \mathbf{A})^{-1} \mathbf{A}^\mathrm{T} \mathbf{N}^{-1} \mathbf{d},
\end{equation}
where $\mathbf{N}$ is the data noise covariance matrix,
$\mathbf{x}^\mathrm{T}$ is the transpose and $\mathbf{x}^{-1}$ is the
inverse of matrix $\mathbf{x}$. See, e.g., \citet{1997ApJ...480L..87T}
for more details.

\subsection{Solutions to the Map-Making Equation}

The noise covariance matrix encodes noise correlations between all
samples, both in time and between detectors. It is a very large
matrix, $N_\mathrm{d} \times N_\mathrm{d}$, where $N_\mathrm{d}$ is
the product of the number of detectors and number of time
samples. When taking into account the full data covariance matrix,
Equation~\ref{eqn:mmsoln} is not directly solvable for even
moderate-sized data sets. However, several approaches exist for
approximating the full solution.

\altsubsubsection*{Simple Re-binning} 

If the noise correlations can be ignored (or high-pass filtered so
that correlations are insignificant), $\mathbf{N}$ is diagonal and the
map-making equation becomes a weighted mean of all samples that fall
in each pixel. This is also known as ``naive'' map-making.

\altsubsubsection*{Direct Solutions} 

The noise covariance matrix is not directly invertible, but
approximations can be made so that the map-making equation can still
be solved for certain data sets
(e.g.\ \citealt{2008ApJ...681..708P,2013ApJ...762...10D}). These
methods work well for instruments with order 100--1000 detectors, but
have yet to be successfully applied to larger data sets.

\altsubsubsection*{Iterative Solutions} 

An approach more tractable for large data sets is to iteratively solve
for noise models (e.g.\ common mode, atmosphere, $1/f$ noise) along
with the sky model. Current ground-based large-format submillimeter
instruments which use iterative map-makers include: CRUSH for SHARC-2
\citep{2008SPIE.7020E..45K}; BoA for LABOCA
\citep{2012SPIE.8452E..1TS}; and SMURF for SCUBA-2
\citep{2013MNRAS.430.2545C}. The iterative model used by SMURF is
illustrated in Figure~\ref{fig:iterative_flowchart}. A key feature of
the map-maker is that it is modular, allowing it to adapt to the
particular instrument and data set at hand.

\section{Scaling SMURF/SCUBA-2 to Next-Generation Instruments}

The next generation of ground-based submm instruments \citep[such as
CCAT:][]{P10_adassxxiii}  will make use of Kinetic
Inductance Detectors (KIDs; e.g.\ \citealt{2003Natur.425..817D}),
allowing for 10,000s of detectors in the focal plane. As a concrete
example, we consider an instrument with 30,000 detectors sampled at 2
kHz. The data volume scaling from SCUBA-2 to the example Next-Gen
instrument is listed in Table~\ref{tab:scaling}.
\begin{table}[t]
\caption{Data volume scaling from SCUBA-2 to Next-Gen Instrument\label{tab:scaling}}
\centering
\begin{tabular}{lcc}
\hline \hline
& \textbf{SCUBA-2} & \textbf{Next Gen} \\ \hline
No.\ of detectors & 5120 & 30,000 \\
Data rate & 170\,Hz & 2000\,Hz \\
No.\ of samples in 15 min & $0.8\times10^9$ & $54\times10^9$ \\
Memory to store data & 6\,GB & 430\,GB \\
Scaling factor & 1 & 70 \\ \hline \hline
\end{tabular}
\end{table}
A SMURF reduction of 16 minutes of SCUBA-2 450\,\micron\ data takes
about 7 minutes on a modern server with 8 cores at 2.67 GHz, using
33\,GB of memory.\footnote{Note that this includes memory for the
  noise models, some using memory equal to the data size, and is thus
  several times larger than the memory needed to store the raw
  detector data.}  Assuming linear scaling (non-linear components are
sub-dominant), reducing the Next-Gen data volume on the same machine
would require 2.3\,TB of memory and would take 8 hours. In five years,
we might imagine a 32-core machine with 2\,TB memory, and modest
improvements in CPU speed. Such a machine can reduce the 15 minutes of
Next-Gen data in about an hour.

\begin{figure}[t]
\centering
\includegraphics[width=0.6\textwidth]{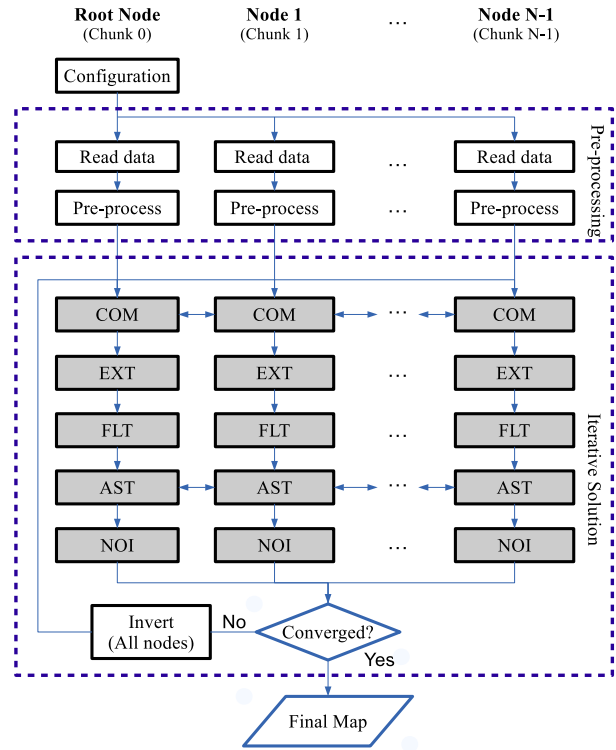}
\caption{A schematic for how distributed-memory parallelization of
  SMURF might work. Horizontal arrows indicate modules where
  inter-node communication is required. The others are trivially
  parallelized.\label{fig:parallel_flowchart}}
\end{figure}

To further decrease runtime, distributed-memory parallel processing
will be required. SMURF is written to take advantage of multi-core
systems, using a shared-memory model; certain modules are trivially
distributed, while others will require inter-node communication. See
Figure~\ref{fig:parallel_flowchart}.

\section{Conclusions}

Rigorous map-making inversions will be infeasible for next generation
instruments, so some approximate solution, using iterative methods,
will be necessary. Scaling to a real example of SCUBA-2 data reduction
by SMURF, we see that reducing 15 minutes of data on a single machine
should be possible using existing software. These individual maps can
be co-added for longer observations. Should reducing the data in 15
minute chunks not be sufficient for recovering the angular scales of
interest, distributed-memory parallelization should be possible.

\bibliography{P14}

\end{document}